\documentclass[aps,prl,floatfix,twocolumn,groupedaddress,showpacs,letterpaper,amsmath,amssymb]{revtex4}
\usepackage{amsfonts,mathrsfs,amsthm}
\usepackage{amsmath,amssymb,graphicx,hyperref}
\hypersetup{pdfnewwindow=true, colorlinks=true, linkcolor=blue, anchorcolor=blue, citecolor=blue, filecolor=blue, menucolor=blue, urlcolor=blue}
\newcommand{\beq}{\begin{equation}}
\newcommand{\eeq}{\end{equation}}
\newcommand{\bea}{\begin{eqnarray}}
\newcommand{\eea}{\end{eqnarray}}

\newcommand{\nn}{\nonumber}
\newcommand{\down}{\downarrow}
\newcommand{\up}{\uparrow}

\begin{document}

\title{One-and-a-half-channel Kondo model and its family of non-Fermi liquids}

\author{Anna I.\ T\'oth}

\affiliation{Bajza utca 50., H-1062 Budapest, Hungary}

\date{\today}

\begin{abstract}
I construct non-Fermi liquid (NFL) quantum impurity models that are similar to the overscreened
multi-channel Kondo models with the difference that an odd number of electron species screen the impurity. 
The simplest of them, named sesqui-channel (i.e.\ one-and-a-half-channel) Kondo (1.5CK) model, has
less degrees of freedom and is simpler than the two-channel Kondo model, and yet it exhibits NFL physics.
Using representation theory I derive the 1.5CK model for a spin-half impurity surrounded with electrons
in cubic crystal field and solve it with the numerical renormalization group.  
\end{abstract}


\maketitle

\begin{figure}
\includegraphics[width=1.\linewidth]{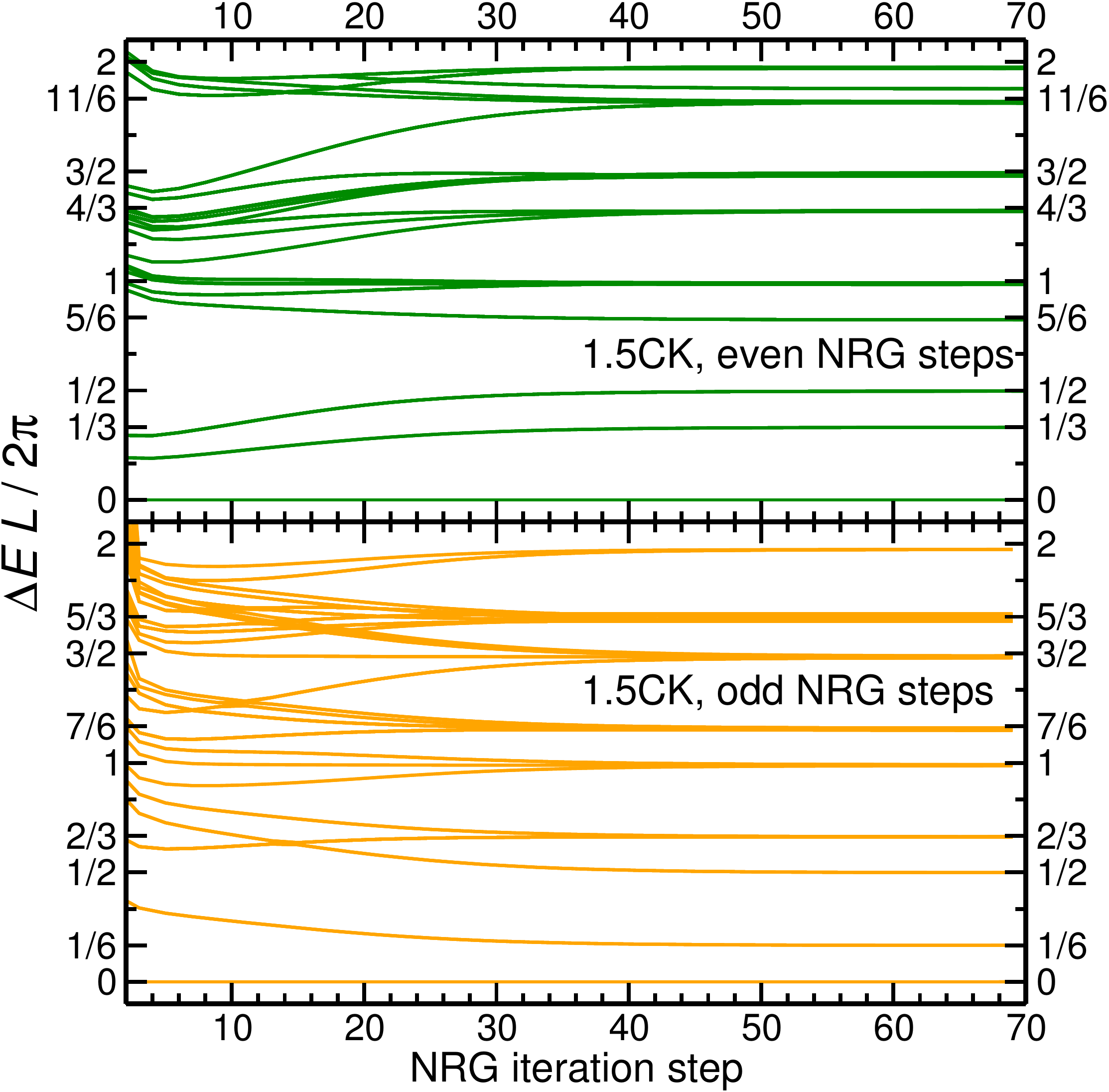}
\includegraphics[width=1.\linewidth]{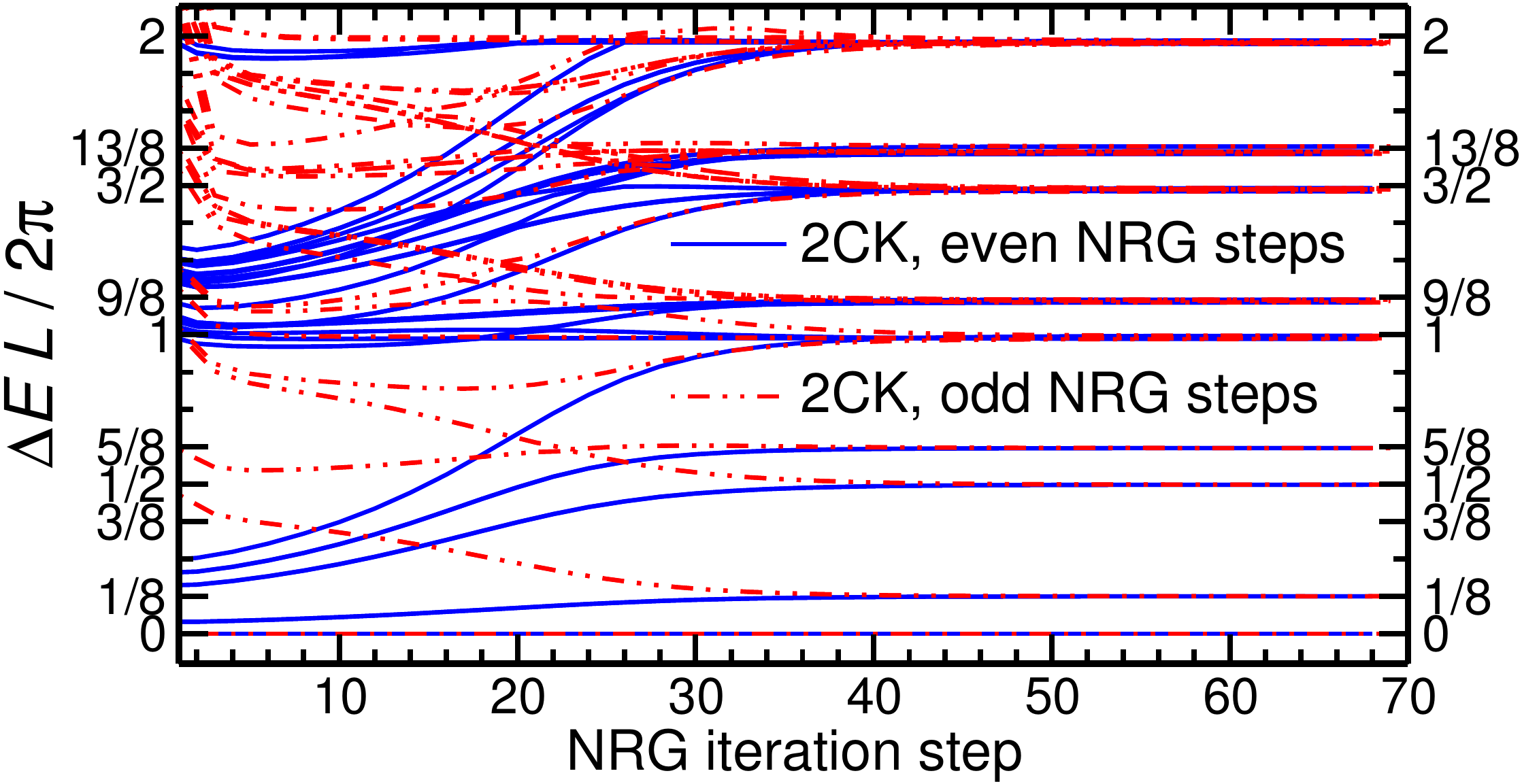}
\caption{\label{fig:spectrum}NRG flow and finite size excitation spectrum of the 1.5CK model (top)
in comparison with the 2CK fixed point spectrum (bottom).  $\Delta E$ stands for the excitation energy of the allowed energy levels and 
$L$ is the system size. The 1.5CK fixed point shows even-odd oscillation with the changing parity of the NRG iteration steps,
whereas there is non such oscillation for the 2CK fixed point, and therefore the energy levels corresponding to 
iteration steps with differing parity could have been superimposed on each other without obscuring the figure.
The ratio of the two distinct neighboring level spacings seen in the fixed point spectrums is \,1:2\, for the 1.5CK model, whereas it is
\,1:3\, in the 2CK model and \,1:1\, in a Fermi liquid. (top) ${\cal  H}_\perp^{1.5\textrm{C}\textrm{K}}+{\cal  H}_z^{1.5\textrm{C}\textrm{K}}$ was solved
at ${\cal J}=0.25 D$ (in units of the bandwidth, $D$) with discretization parameter $\Lambda=2$ keeping at most 1000 states of the U(1) charge symmetry
at each NRG step. (bottom) 2CK fixed point spectrum using the SU$(2)_{\textrm spin}\times$SU$(2)_{\textrm charge 1}\times$SU$(2)_{\textrm charge 2}$ symmetry of
the model while keeping at most 500 multiplets at each NRG step.}  
\end{figure}

\paragraph{}
Identifying the microscopic origin of non-Fermi liquid (NFL) types of metallic behavior, in most cases remains an outstanding 
challenge, even though many years have passed since the discovery of NFL phenomena in heavy fermions \cite{Ott83,Seaman91} and 
in high-$T_c$ cuprate superconductors (SC's) \cite{Suzuki87,Gurvitch87,Ginsbergetc}. 
More recent encounters with NFL physics took place in high-$T_c$ iron-pnictide   
\cite{Liu08,Gooch09,Fang09,Kasahara10} and in certain iron-chalcogenide SC's \cite{Stojilovic10}.
It has been found in stoichiometric transition metal oxides such as VO$_2$ \cite{Allen93,Qazilbash06}
and some ruthenates \cite{Kostic98,Khalifah01,Lee02} 
and iridates \cite{Nakatsuji06,Cao07}, as well as in intermetallics \cite{Pfleiderer01}
and pure transition metal compounds \cite{Steiner03,Brando08}
and other $d$- and $f$-electron systems \cite{Stewart} and elsewhere \cite{Dardel93,Moser98,Bockrath99}. 
Disadherence to Fermi liquid theory has many forms. 
In heavy fermions it manifests itself, among others, in 
the electronic specific heat (or Sommerfeld) coefficient, the magnetic susceptibility  and 
the electrical resistivity, which show either
diverging/logarithmic or mild power-law $T$-dependencies down to the lowest temperatures attainable 
\cite{Stewart}.
The phase diagram of hole-doped cuprates based on resistivity \cite{Barisic13} 
and neutron scattering \cite{Li08} measurements has only recently been redrawn along with universality asserted 
\cite{Barisic13}. Then it was observed that the pseudogap 
part of their metallic phase exhibits Fermi liquid (FL) properties \cite{Mirzaei13,Chan14}.
Nevertheless, 
around optimal doping for an extended region above the pseudogap and superconducting transition temperatures 
their in-plane transport properties do not fit in with FL theory: 
the resistivity is linear with $T$ \cite{Suzuki87,Ando04}, 
the Hall coefficient has a pronounced $T$-dependence \cite{Chien91}, 
and the frequency dependence of the in-plane optical conductivity \cite{Schlesinger90} is also out of the FL realm.
Further NFL features are seen 
in the electronic Raman spectra \cite{Blumberg98} and in nuclear magnetic resonance studies \cite{Berthier96}.
The range of NFL physics around optimal doping extends to lower temperatures when superconductivity is quenched by a magnetic field. 
Tested this way, the violation of the Wiedemann--Franz law \cite{Hill01} also points to a NFL ground state.
The anomalous normal state transport properties of iron-pnictides are similar to those of the cuprates \cite{Harris95,Kasahara10}.
Linear-in-$T$ resistivity has further been observed in some of the heavy fermion compounds, 
in VO$_2$, and in certain ruthenates \cite{Grigera01,Khalifah01}, 
whereas in other heavy fermions and ruthenates
other functional forms of $T$, but still clear deviations from FL behavior are 
displayed in the resistivity and in the optical conductivity.
NFL phenomena discovered in three- and low-dimensional materials are abundant and 
diverse,
whereas 
there is only a handful of  well-established NFL models and principles leading to NFL behavior.  
They include  on the phenomenological level
the marginal Fermi liquid theory \cite{Varma89} 
which has been invoked to successfully describe 
many of 
the anomalous normal state properties of cuprates. 
NFL physics can also appear as a consequence of disorder \cite{Miranda05} as e.g.\ in doped semiconductors, disordered heavy fermion 
systems, or metallic glass phases.  
Two major classes of microscopic models that became paradigms for describing NFL physics are 
the Tomonaga--Luttinger liquid (TLL) \cite{Tomonaga50_Luttinger63} and the overscreened multi-channel Kondo models \cite{Nozieres80}.
The former one accounts for the peculiarities of a broad class of one-dimensional metals. 
It features  spin-charge separation, i.e.\ its spin and charge degrees of freedom propagate with different velocities. 
Consequently, although it has coherent low-energy excitations, their quantum numbers differ from those of the electron, 
which makes the TLL alike the overscreened multi-channel Kondo models.  
These latter ones are quantum impurity models where a spin degree of freedom couples to several degenerate baths of 
non-interacting electrons which correspond to the screening channels.
The simplest variant of overscreened Kondo models was considered to be the spin-half two-channel Kondo (2CK) model introduced in 1980
\cite{Nozieres80,Zawadowski80}.  
To advance the microscopic understanding of NFL phenomena,
in this Letter I construct and study new types of quantum impurity models with new
types of NFL behavior that have not been identified before. 
The simplest NFL quantum impurity model I introduce, which I named sesqui-channel (i.e.\ one-and-a-half-channel) Kondo (1.5CK) model,
has less degrees of freedom 
than the two-channel Kondo (2CK) model, and yet it still exhibits NFL physics.
I solve the 1.5CK model using the numerical renormalization group (NRG) \cite{Wilson75}.
\paragraph{Construction of the 1.5CK and other half-integer-channel, overscreened Kondo-type models.}
In Ref.\  \cite{Nozieres80},  Nozi\`eres and Blandin addressed the behavior of magnetic impurities in metals. 
They analyzed an Anderson-type of Hamiltonian taking into account not only the spin but also the orbital 
degrees of freedom of the impurity, the spin-orbit coupling and  crystal field 
effects. Using only symmetry and scaling arguments they were able to make predictions about the 
stability of the strong coupling fixed point of the model.  
While studying the strong coupling fixed point of the orbital singlet case 
in an isotropic environment and switching to an effective Kondo-type of description, they have come to 
the conclusion that there exists an intermediate Kondo coupling fixed point 
with a non-trivial ground state for the special case of $n > 2S$, with $n$ the number of conduction electron screening channels and 
$S$ the value of the impurity spin.  Kondo models satisfying the criterion $n > 2S$ are called  overscreened Kondo models.
The prediction about their NFL fixed point got its first numerical confirmation from Cragg and Lloyd's 
NRG calculations in the case of $n=2$ and $S=1/2$ \cite{Cragg80}.
The corresponding model is  the spin-half two-channel Kondo (2CK) model which has so far been considered to realize
the simplest case of overscreening.
The case in which $n=1.5$ and $S=1/2$, i.e.\ the number of screening channels is a half-integer, 
has not been looked at though.  
At first sight this choice might seem peculiar, but what the number $n$ really stands for is half 
the number of electron species that screen the spin. 
One might envision this case as e.g.\ an impurity with a Kramers doublet ground state  in a local environment with  cubic symmetry, 
with surrounding $T_{2g}$ conduction electrons whose 
spin degeneracy is lifted.  
To construct a NFL impurity model out of this configuration one can proceed by observing that the 
ground state of the one-channel Kondo model can be guessed simply by diagonalizing the local Hamiltonian, whereas in case of the 
two-channel Kondo model the same process does not work.  The additional channel  introduces frustration.  Thus when 
constructing the local part of the 1.5CK model, 
characterized by $S=1/2$, $n=1.5$ and a NFL fixed point,  
I introduce a similar kind of frustration but 
with only three flavors of conduction electrons. In the first step I construct the spin-flip part of the 1.5CK Hamiltonian,
${\cal  H}_\perp^{1.5\textrm{C}\textrm{K}}$, as  diagonal processes
in themselves cannot bring about NFL behavior.
\bea\label{eq:spinflipHam}
{\cal  H}_\perp^{1.5\textrm{C}\textrm{K}}&\equiv&
{\cal J}_{\perp}^{\,ab\,}S^+_{} a^\dagger_{} b^{}_{}\,+\,{\cal J}_{\perp}^{\,bc\,}S^+_{} b^\dagger_{} c^{}_{}\,+\,h.c.
\eea
where  $a^\dagger,b^\dagger,c^\dagger$ create the three different species of conduction electrons at the site of the spin-half impurity, $\vec S$,
and $S^+\equiv S^x+iS^y$. When ${\cal J}_{\perp}^{\,ab\,}={\cal J}_{\perp}^{\,bc\,}$ this Hamiltonian together with a term that accounts for the kinetic
energy of the conduction electrons has a NFL fixed point.
The structure of its excitation spectrum is shown
in Fig.\ \ref{fig:spectrum} next to the well-known 2CK fixed point spectrum for easy comparison.
Eq.\ \eqref{eq:spinflipHam} can be complemented to gain the form
\bea
{\cal  H}_\perp^{\,chain,1.5\textrm{C}}&\equiv&
{\cal J}_{\perp}^{}S^+_{} \left(a^\dagger_{} b^{}_{}\,+\,b^\dagger_{} c^{}_{}\,+\,c^\dagger_{} a^{}_{}\right)+\,h.c.\nonumber
\eea
The ground state of this Hamiltonian can be found easily as it leads to FL physics. By adding one more flavor of
conduction electron the 2CK fixed point can be obtained by leaving out one or more arbitrary term(s),
from ${\cal  H}_\perp^{\,chain,2\textrm{C}}$ while keeping  all the 
four different conduction electron species, with creation operators $\psi_i^\dagger\, (i=1,\dots,4)$, in the spin-flip Hamiltonian
\begin{multline}
{\cal  H}_\perp^{\,chain,2\textrm{C}}\,\equiv\,
{\cal J}_{\perp}^{}S^+_{} \left(\psi^\dagger_{1} \psi_2^{}\,+\,\psi_2^\dagger \psi_3^{}\,+\,\psi_3^\dagger \psi_4^{}+\,\psi_4^\dagger
\psi_1^{}\right)\\\,+\,h.c.\nonumber
\end{multline}
I conjecture that for larger number of conduction electron species the construction goes the same way, i.e.\ the spin-flip part of a NFL, overscreened Kondo-type Hamiltonian for $n=2k+1$ conduction electron species can be written e.g.\ as
\bea
    {\cal  H}_\perp^{\,k.5\textrm{C}\textrm{K}}&\equiv&
    {\cal J}_{\perp}^{}S^+_{} \left(\psi^\dagger_{1} \psi_2^{}\,+\,\dots\,+\,\psi^\dagger_{n-1} \psi_n^{}\right)\,+\,h.c.,\nonumber
\eea
and presumably has a ratio of $1:(n-1)$
between the two different level spacings in the corresponding NFL fixed point spectrum.
As dictated by universality the fixed point properties depend only on the value of
the impurity spin and on the number of conduction electron screening channels, i.e.\ on the number of degrees of freedom of the model.
I expect that the same process works for $n=2k$ as well, and produces spin-flip Hamiltonians leading to the well-known overscreened multi-channel Kondo fixed points. I did not confirm the above conjectures with NRG due to their higher computational cost and the less likelihood for the experimental
realization of higher channel-number NFL Kondo physics.

Using representation theory the 1.5CK Hamiltonian can also be derived as the one relevant coupling
between a spin-half impurity in a cubic field surrounded with $T_{2g}$ conduction electrons whose spin
degeneracy is lifted.  The derivation is similar in spirit to Cox's derivation \cite{Cox87} with the difference
that he was looking for a natural symmetry arrangement for 2CK physics to emerge.  In the derivation I use the notation of ref.\ \cite{Koster63}
when referring to point groups and
their irreducible representations (irreps).  The first step is to find a point group which has a three-dimensional (3D) irrep.
One such group is $T_d$ of all proper rotations sending a regular tetrahedron into itself.
The derivation goes the same way and results in the same Hamiltonian for other groups, $T$,
$T_h$ and $O$ with 3D irreps as the Clebsh--Gordan coefficients are identical  \cite{Koster63}. I choose the impurity
electron to be the $\Gamma_6$ irrep of $T_d$ which transforms as a spin-half spinor under $T_d$.
I construct all possible electron-hole operators out of this irrep
based on the tensor product $\Gamma_6\otimes\Gamma_6\,=\Gamma_1\oplus\Gamma_4$. The irrep $\Gamma_1$ (or $A_1$) is the trivial representation of $T_d$,
whereas $\Gamma_4$ is 3D and transforms as the three components of the spin operator ($S_x,S_y,S_z$) under $T_d$.
The only irreducible $\Gamma_4$ tensor operator that can be composed from the two $\Gamma_6$ spinors, 
$\,
\left[
\begin{array}{c}
f^\dagger_{\up}\\
f^\dagger_{\down}
\end{array}
\right],\,
\left[
\begin{array}{c}
-\,f^{}_{\down}\\
f^{}_{\up}
\end{array}
\right],\,$
with $f^\dagger_{\mu}$ creating a spin-half impurity with spin $\mu$, and normalized to anticommute as
$\left\{f^\dagger_{\mu},f^{}_{\nu}\right\}=\delta_{\mu,\nu}$, with $\mu,\nu\in\left\{\up,\down\right\}$,  
 is
\bea
\frac{1}{ \sqrt 2}\left[
\begin{array}{c}
f^\dagger_{\down}f^{}_{\up}+f^\dagger_{\up}f^{}_{\down}\\
i\left(f^\dagger_{\down}f^{}_{\up}-f^\dagger_{\up}f^{}_{\down}\right)\\
f^\dagger_{\up}f^{}_{\up}-f^\dagger_{\down}f^{}_{\down}
\end{array}
\right].\nonumber
\eea
As for the conduction electron part of the local Hamiltonian,
from the two $\Gamma_5$ conduction electron and hole tensor operators,
$\,
\left[
\begin{array}{c}
c^\dagger_{x}\\
c^\dagger_{y}\\
c^\dagger_{z}\\
\end{array}
\right],\,$ and
$\left[
\begin{array}{c}
  c_x\\
  c_y\\
c_z
\end{array}
\right],\,$ respectively,
whose spin degeneracy is lifted,
one irreducible, electron-hole, $\Gamma_4$
tensor operator can be  composed 
according to the rule
$\Gamma_4\otimes\Gamma_4\,=\Gamma_1\oplus\Gamma_3\oplus\Gamma_4\oplus\Gamma_5$.
This operator is 
\bea
\frac{i}{\sqrt 2}\left[
\begin{array}{c}
  c^\dagger_yc_z-c^\dagger_{z}c_y\\
c^\dagger_{z}c_x-c^\dagger_{x}c_z\\
  c^\dagger_{x}c_y-c^\dagger_{y}c_z
\end{array}
\right]\nonumber
\eea
using the appropriate Clebsch--Gordan coefficients.
Thus the non-trivial (i.e.\ non-potential scattering) spin-flip part of the resulting local Hamiltonian is
\begin{multline}
  {\cal  H}_\perp^{1.5\textrm{C}\textrm{K}}\,\equiv\,{\cal J}\left[
    \frac{i}{2}\left(f^\dagger_\up f^{}_\down +f^\dagger_\down f^{}_\up\right)\left(c^\dagger_y c^{}_z -c^\dagger_z c^{}_y\right)\,+\right.\\
    \left.+\,\frac{1}{2}\left(f^\dagger_\up f^{}_\down -f^\dagger_\down f^{}_\up\right)\left(c^\dagger_z c^{}_x -c^\dagger_x c^{}_z\right)\right],\nn
\end{multline}
whereas the diagonal part becomes
\bea
    {\cal  H}_z^{1.5\textrm{C}\textrm{K}}\,\equiv\,{\cal J}\,\frac{i}{2}\left(f^\dagger_\up f^{}_\up -f^\dagger_\down f^{}_\down\right)
    \left(c^\dagger_x c^{}_y -c^\dagger_y c^{}_x\right)\,.\nn
\eea
Thus with the identifications $a^\dagger\equiv\left(-c^\dagger_x+ic^\dagger_y\right)/\sqrt{2}$, $b^\dagger\equiv c_z^\dagger$,
$c^\dagger\equiv\left(c^\dagger_x+ic^\dagger_y\right)/\sqrt{2}$, 
we regain the form stated in Eq.\ \eqref{eq:spinflipHam} plus a diagonal supplement, ${\cal  H}_z^{1.5\textrm{C}\textrm{K}}$,
reading
\bea
{\cal  H}_\perp^{1.5\textrm{C}\textrm{K}}\,\equiv\,{\cal J}\,\frac{1}{\sqrt{2}}\,f^\dagger_\up f^{}_\down\left(a^\dagger b^{}+b^\dagger c^{}\right)\,+\,h.c.\,,\\
{\cal  H}_z^{1.5\textrm{C}\textrm{K}}\,\equiv\,{\cal J}\,\frac{1}{2}\left(f^\dagger_\up f^{}_\up -f^\dagger_\down f^{}_\down\right)
\left(a^\dagger a^{} -c^\dagger c^{}\right)\,,
\eea
with $f^\dagger_\up f^{}_\down\equiv S^+$. 
\paragraph{Energy spectrum of the 1.5CK model.} To confirm that both ${\cal  H}_\perp^{1.5\textrm{C}\textrm{K}}$ and
${\cal  H}_\perp^{1.5\textrm{C}\textrm{K}}+{\cal  H}_z^{1.5\textrm{C}\textrm{K}}$
together with the kinetic
energy of the conduction electrons indeed flow to a NFL fixed point, I solved them with NRG \cite{Wilson75,Toth08}.
During the calculations I could make use of only the charge symmetry of these models, as they lack spin symmetry.  I obtained the same,
1.5CK fixed point spectrum for both Hamiltonians. The NRG flow and the 1.5CK fixed point spectrum of
${\cal  H}_\perp^{1.5\textrm{C}\textrm{K}}+{\cal  H}_z^{1.5\textrm{C}\textrm{K}}$ for ${\cal J}=0.25$ in units of the bandwidth
are shown in Fig.\ \ref{fig:spectrum} next to the 2CK spectrum. 
The ratio of the two neighboring level spacings in the excitation spectrums is 1:2
for the 1.5CK model, whereas it is 1:3 in the 2CK model and 1:1 in a Fermi liquid. Thus, in
this regard, the 1.5CK model is a more basic NFL quantum impurity model than the 2CK model.
The 1.5CK fixed point spectrum shows even-odd oscillation due to particle-hole symmetry and that 
three more conduction electron creation operators are introduced in each NRG iteration step. That is the same argument that was applied in
Wilson's paper for the one-channel Kondo model \cite{Wilson75}.
As expected, the physical properties computed around the 1.5CK fixed point cannot depend on the parity of the NRG iteration steps.
This assertion was checked numerically for the specific heat.
Presenting the thermodynamic and dynamic properties of the half-integer channel Kondo models using NRG will be the subject of a further study, just as
extending conformal field theoretical (CFT)
considerations \cite{Affleck91} to these models in order to understand their excitation spectrum and other properties.
From e.g.\ the CFT solutions we know that in the 2CK
model the electronic specific heat coefficient and the magnetic susceptibility both diverge as $\log T$ when
$T \to 0$, whereas for higher channel numbers the divergence is faster as $T\to 0$. An 
intriguing question is whether the same considerations apply to the 1.5CK model, and
how the divergence of these quantities is affected as $T \to 0$.  
\paragraph{Conclusion.}
I presented a family of NFL quantum impurity models where a spin-half impurity is exchange coupled to an odd number conduction electron species.
Computing and understanding the physical properties of these models, and  matching them with physical properties observed in materials
are yet to follow.
\paragraph{Acknowledgment.} I thank Silke B\"uhler-Paschen and Karsten Held for many enlightening discussions and 
Zhicheng Zhong for thought-provoking correspondence.

\end{document}